\documentclass{ws-p8-50x6-00}
\usepackage{amssymb}

\newcommand{\beq}[1]{\begin{equation}\label{#1}}
\newcommand{\eeq}{\end{equation}}
\newcommand{\beqa}[1]{\begin{eqnarray}\label{#1}}
\newcommand{\eeqa}{\end{eqnarray}}

\begin{document}

\title{Collective flow and multiparticle azimuthal correlations}

\author{N. BORGHINI}

\address{Service de Physique Th\'eorique, CP225, 
Universit\'e Libre de Bruxelles, \\
1050 Brussels, BELGIUM\\
E-mail: Nicolas.Borghini@ulb.ac.be}

\author{P. M. DINH and J.-Y. OLLITRAULT}

\address{Service de Physique Th\'eorique, CE-Saclay, \\
91191 Gif-sur-Yvette cedex, FRANCE}  

\maketitle

\abstracts{
The measurement of azimuthal distributions in nucleus-nucleus collisions
relies upon the assumption that azimuthal correlations between particles
result solely from their correlation with the reaction plane (i.e. flow). 
We show that at SPS energies, the ansatz is no longer valid, and two-particle 
correlations due to momentum conservation, final state interactions or 
resonance decays become of the same order as those arising from flow. 
This leads us to introduce new methods to analyse collective flow, based on a 
cumulant expansion, which enable to extract smaller values than those 
accessible to the standard analysis.}

In a collision between two heavy ions, the azimuthal distribution of
outgoing particles with respect to the reaction plane, i.e. collective flow,
is expected to reveal new insights on the central region of the collision:
thermal equilibrium, equation of state, time evolution.\cite{flow-interest} 
It is therefore important to have reliable flow values, and to be able to
measure small values. 

We first discuss in Sec.~\ref{s:standard} the method usually used to analyze
flow data, and we point out that it is based on an assumption which is not
valid at ultrarelativistic energies. 
In the next two sections, we introduce a new method which allows flow
measurements at such energies: the main ideas are presented in
Sec.~\ref{s:multi}, while the practical implementation in terms of the event
flow vector, including acceptance corrections, is given in
Sec.~\ref{s:Qvector}. 
Finally, we briefly summarize our results in Sec.~\ref{s:summary}.

\section{Flow and two-particle correlations}
\label{s:standard}

The standard method for analyzing the experimental data relies on a study of
two-particle azimuthal correlations.\cite{poskanzer98} 
The one-particle azimuthal distribution is first expanded into a Fourier
series, whose coefficients $v_n$ characterize the flow:\cite{voloshin96} 
\beq{vn}
v_n = \left\langle e^{in\phi} \right\rangle,
\eeq
where $\phi$ denotes the azimuth with respect to the reaction plane, while
the average is performed over a large number of events. 

Since the actual orientation of the reaction plane is not known, only
differences between particle azimuths can be measured. 
Thus, flow is extracted from the measured two-particle correlations $\langle
e^{in(\phi_1-\phi_2)} \rangle$, under the assumption that they are only due
to flow, or at least that the contribution $\langle e^{in(\phi_1-\phi_2)}
\rangle_c$ of direct, nonflow correlations is negligible. 
In other words, it is assumed that in the decomposition
\beq{c2}
\left\langle e^{in(\phi_1-\phi_2)} \right\rangle = v_n^2 + \left\langle
e^{in(\phi_1-\phi_2)} \right\rangle_{\! c} 
\eeq
the second term in the right-hand side is much smaller than the first.

However, we have shown that this ansatz is not valid at SPS
energies:\cite{nous1&2} nonflow two-particle correlations, which can stem
from various sources, are of the same magnitude as two-particle correlations
due to flow. 
Thus, the NA49 values for pion and proton flow are significantly modified
when two-particle azimuthal correlations arising from momentum conservation,
HBT quantum effects, and resonance decays are subtracted from the measured
correlations. 

Two-particle correlations may not be easy to take into account in the
standard analysis, even if the sources are well identified (as for example
in the case of resonance decays). 
Moreover, the possibility that there are still unknown sources cannot be
discarded. 
Thus, the safest approach consists in using a method which does not require
a precise knowledge of these correlations.

\section{Measuring flow using multiparticle azimuthal correlations}
\label{s:multi}

We recently proposed a new method for the analysis of flow, which does not
depend on two-particle correlations since these are eliminated by means of a
cumulant expansion of multiparticle azimuthal correlations.\cite{nous3} 
In this section, we show how this method allows measurements of integrated
flow (i.e.\ averaged over a phase space region, see \ref{s:int1}) as well as
differential flow for given transverse momentum and rapidity (\ref{s:diff1}).

\subsection{Integrated flow}
\label{s:int1}

The main idea underlying our method is the following. 
In a collision where $N$ particles are emitted, {\em direct} $k$-particle
correlations are typically of order $1/N^{k-1}$, so that they become smaller
when $k$ increases. 
In particular, their magnitude decreases faster than the flow contribution
to the {\em measured} $k$-particle correlations: one can expect that the
comparison between both terms allows measurements of smaller and smaller
flow values. 

Let us consider for instance the measured four-particle azimuthal
correlations. 
These correlations can be expanded as the sum of products of direct one-,
two- and three-particle correlations, where ``one-particle correlation''
means correlation with the reaction plane, that is flow. 
The order of magnitude of each term in the expansion can easily be
estimated, keeping in mind that a flow term is of order $v_n$ while direct
$k$-particle correlations are $O(1/N^{k-1})$.\cite{nous3} 
After calculation, the dominant terms in the expansion are: 
\beqa{c4}
\left\langle e^{in(\phi_1+\phi_2-\phi_3-\phi_4)}\right\rangle & \simeq &
v_n^4 + \left\langle e^{in(\phi_1-\phi_3)}\right\rangle_{\! c}
\left\langle e^{in(\phi_2-\phi_4)}\right\rangle_{\! c} \\
 & & +\left\langle e^{in(\phi_1-\phi_4)}\right\rangle_{\! c}
\left\langle e^{in(\phi_2-\phi_3)}\right\rangle_{\! c}
+\left\langle e^{in(\phi_1+\phi_2-\phi_3-\phi_4)}\right\rangle_{\! c}.
\nonumber
\eeqa
[For simplicity, we have assumed that $v_n$ is not much smaller than
$v_{2n}$, and neglected a term $O(v_{2n}^2/N^2)$.] 
The first and the fourth terms in r.-h.~s.\ are of magnitude $v_n^4$ and
$1/N^3$ respectively: a direct comparison between them would allow flow
measurements down to values $v_n \gg 1/N^{3/4}$, smaller than the $v_n \gg
1/N^{1/2}$ limit of the standard analysis which can be deduced from
Eq.~(\ref{c2}). 
In order to compare these two terms, we have to get rid of the second and
third terms, which are in fact equal, and correspond to two-particle
correlations. 

That can be done, taking the cumulant of the four-particle azimuthal
correlation. 
The purpose of this cumulant is precisely to remove lower order (i.e.\
involving $k'$ particles with $k' < k = 4$) correlations. 
It is defined as the difference between measured correlations $\langle
e^{in(\phi_1+\phi_2-\phi_3-\phi_4)}\rangle - 2 \langle
e^{in(\phi_1-\phi_3)}\rangle^2$. 
Using Eqs.~(\ref{c2}) and (\ref{c4}), one finds that the cumulant is:
\beq{c4c}
\left\langle\!\!\left\langle
e^{in(\phi_1+\phi_2-\phi_3-\phi_4)}\right\rangle\!\!\right\rangle = 
-v_n^4 + \left\langle e^{in(\phi_1+\phi_2-\phi_3-\phi_4)}\right\rangle_{\!
c} = 
-v_n^4 + O\left(\frac{1}{N^3}\right),
\eeq
so that it gives the flow, provided this latter is larger than $1/N^{3/4}$. 
Measuring this cumulant is indeed sensitive to smaller flow values than the
standard method. 
For instance, at SPS energies where about $N = 2500$ particles are emitted,
using the cumulant (\ref{c4c}) allows measurements of $v_n \gg 0.3\%$, while
the standard method is limited to $v_n \gg 2\%$; these values are to be
compared with the published values $v_1 \simeq v_2 \simeq 3\%$ for
pions.\cite{NA49flow}

\subsection{Differential flow}
\label{s:diff1}

Let us assume that the average value $v_n = \langle e^{in\phi} \rangle$ of
flow is known for a given particle species, which we shall call ``pions'' to
fix ideas. 
For any type of particles, let us say ``protons'', although it can be
anything---even ``pions''---, it is possible to perform detailed
measurements of the flow $v'_m(p_T,y) = \langle e^{im\psi} \rangle$ using
multiparticle correlations in order to go beyond the limitations of the
standard two-particle method.\cite{nous3} 
To simplify the expressions, we shall assume $n=m$; however, the ideas
remain valid is $m$ is a multiple of $n$. 

Let us consider for example the four-particle azimuthal correlation between
a proton and three pions $\langle e^{in(\psi+\phi_1-\phi_2-\phi_3)} \rangle$.
As above, this measured correlation can be expanded in terms of lower order
direct correlations, yielding a term $v'_n v_n^3$, a term of order $1/N^3$
corresponding to direct four-particle correlations, and other terms.
These latter, and in particular two-particle correlations, can be removed
taking the cumulant [by analogy with Eq.~(\ref{c4c})]
\beqa{c4c-diff}
\left\langle\!\!\left\langle
e^{in(\psi+\phi_1-\phi_2-\phi_3)}\right\rangle\!\!\right\rangle & \equiv &
\left\langle e^{in(\psi+\phi_1-\phi_2-\phi_3)} \right\rangle - 2
\left\langle e^{in(\psi-\phi_2)} \right\rangle  \left\langle
e^{in(\phi_1-\phi_3)} \right\rangle \\    
& \simeq &  -v'_n v_n^3 + \left\langle
e^{in(\psi+\phi_1-\phi_2-\phi_3)}\right\rangle_{\! c} =  
-v'_n v_n^3 + O\left(\frac{1}{N^3}\right). \nonumber
\eeqa
Inspection of this equation reveals that the measurement of the cumulant in
the left-hand side gives access to $v'_n$ as soon as it is larger than
$1/(Nv_n)^3$, which is also the accuracy on $v'_n$.

\section{Practical implementation using the event flow vector}
\label{s:Qvector}

\subsection{Integrated flow}
\label{s:int2}

An easy way to implement the method outlined in section~\ref{s:multi}
consists in using the event flow vector, which is defined as 
\beq{Qn}
Q_n = \frac{1}{\sqrt{M}} \sum_{j=1}^M e^{in\phi_j},
\eeq
where the sum runs over particles from the same event, and $M$ should be
chosen as large as possible. 
Averaging Eq.~(\ref{Qn}) over many events, one finds $\langle Q_n \rangle =
\sqrt{M} v_n$: a nonvanishing $\langle Q_n \rangle$ signals collective flow. 

From definition (\ref{Qn}), it is obvious that the powers of $|Q_n|^2$
involve multiparticle azimuthal correlations, and only differences between
angles (which require no knowledge of the reaction plane orientation). 
That explains why the method can naturally be expressed in terms of the flow
vector: the average value $\langle Q_n \rangle$ corresponds to $v_n$, i.e.\
what we want to extract from the data, while the measured quantity $\langle
|Q_n|^k \rangle$ corresponds to measured $k$-particle azimuthal correlations. 

Let us turn once more to four-particle correlations. 
By analogy with what was found in section~\ref{s:multi}, the expansion of
the moment $\langle |Q_n|^4 \rangle$ involves various terms: $\langle Q_n
\rangle^4$, a term $\langle |Q_n|^4 \rangle_c$ corresponding to direct
four-particle correlations, and other terms which we wish to get rid of. 
These latter can indeed be removed, using the cumulant of the $Q_n$
distribution, which at order $k = 4$ is defined as 
\beq{Q4c}
\left\langle\!\left\langle |Q_n|^4 \right\rangle\!\right\rangle \equiv 
\langle |Q_n|^4 \rangle - 2 \langle |Q_n|^2 \rangle^2 = 
-\langle Q_n \rangle^4 + O\left(\frac{1}{M}\right).
\eeq
Since $\langle Q_n \rangle = \sqrt{M} v_n$, this equation is strictly
equivalent to Eq.~(\ref{c4c}), provided $M$ and $N$ are of the same
magnitude. 
Therefore, the measurement of the cumulant (\ref{Q4c}) provides a way to
extract smaller $v_n$ values than what can be obtained within the standard
two-particle analysis.

\subsection{Acceptance corrections}
\label{s:acceptance}

An important feature of the method is the easy implementation of acceptance
corrections.\cite{nous3} 
The only modification regards the definition of the cumulant: in
Eq.~(\ref{Q4c}), the cumulant involves only two moments of the $Q_n$
distribution because the other terms vanish for a perfectly isotropic
detector. 
If the detector is not isotropic, these terms should be reintroduced in the
definition, and the flow vector $\bar Q_n$ measured in the laboratory frame
is not equivalent to the flow vector $Q_n$ with respect to the reaction
plane. 
Taking into account these changes, the cumulant is (dropping the $n$ for
brevity): 
\beqa{Q4c-a}
\left\langle\!\left\langle |\bar Q|^4\right\rangle\!\right\rangle &
\!\equiv\! &  
\left\langle |\bar Q|^4\right\rangle
-2\left\langle \bar Q\right\rangle\left\langle \bar Q \bar Q^{*
2}\right\rangle 
-2\left\langle \bar Q^*\right\rangle\left\langle \bar Q^*\bar Q^2\right\rangle
-2\left\langle |\bar Q|^2\right\rangle^2
-\left\langle \bar Q^2\right\rangle\left\langle \bar Q^{* 2}\right\rangle \cr
& & + 8\left\langle \bar Q\right\rangle\left\langle \bar
Q^*\right\rangle\left\langle |\bar Q|^2\right\rangle 
+4\left\langle \bar Q\right\rangle^2\left\langle \bar Q^{* 2}\right\rangle
+4\left\langle \bar Q^*\right\rangle^2\left\langle \bar Q^2\right\rangle
-6\left\langle \bar Q\right\rangle^2\left\langle \bar Q^*\right\rangle^2 \cr
& \!=\! &-\left\langle Q\right\rangle^4  + O(1/M).
\eeqa
This expression generalizes Eq.~(\ref{Q4c}) to the case of real, nonperfect
detectors, and should be used to extract flow from experimental data.

\subsection{Differential flow}
\label{s:diff2}

Let us readopt the same notations as in Sec.~\ref{s:int1}.
Representing the pions' azimuths by the flow vector $Q_n$ [Eq.~(\ref{Qn})],
the four-particle (1 proton, 3 pions) correlations considered in
Sec.~\ref{s:int1} appears in the moment $\langle |Q_n|^2 Q_n^* e^{in\psi}
\rangle$. 
The idea is again to expand this moment, and to take the cumulant in order
to remove unwanted terms. 
More precisely, the corresponding cumulant for the (1+3)-particle azimuthal
correlation is: 
\beqa{Q4c-diff}
\left\langle\!\left\langle |Q_n|^2 Q_n^* e^{in\psi}
\right\rangle\!\right\rangle & \equiv & 
\left\langle |Q_n|^2 Q_n^* e^{in\psi} \right\rangle - 2 \left\langle Q_n^*
e^{in\psi} \right\rangle\left\langle |Q_n|^2 \right\rangle \cr 
 & = & - \langle Q_n \rangle^3 v'_n + O\left(\frac{1}{M} \right).
\eeqa
Once the pion flow $\langle Q_n \rangle$ has been measured, this cumulant
gives $v'_n$ with accuracy $O(1/M\langle Q_n \rangle^3) = O(1/(Mv_n)^3)$, as
did the cumulant (\ref{c4c-diff}).

\section{Summary}
\label{s:summary}

We have shown that the standard flow analysis, which relies on two-particle
azimuthal correlations, is in trouble when $v_n$ becomes of order
$1/N^{1/2}$. 
In opposition, the new method we advocate, based on a cumulant expansion of
four-particle correlations, allows integrated flow measurements down to
values $v_n \gtrsim 1/N^{3/4}$. 
It is even possible to extract smaller values, using higher order cumulants,
as for example the cumulant of the six-particle azimuthal correlation, or
equivalently the cumulant $\langle\!\langle |Q_n|^6 \rangle\!\rangle$: we
have derived a generating equation which gives the cumulant of the $Q_n$
distribution at a given order, and relates it to the corresponding power of
$\langle Q_n \rangle$.\cite{nous3} 

This new method also allows the measurement of differential flow, from the
correlation between a particle azimuth and the event flow vector $Q_n$. 
Once again, the results of Sec.~\ref{s:diff2} can be generalized to
arbitrary order, using a generating equation.\cite{nous3}. 
Finally, corrections for detector inefficiencies are naturally implemented
in the method, using a redefinition of the cumulants. 

We wish to emphasize that all the cumulants, either of multiparticle
azimuthal correlations or of the $Q_n$ vector, can be obtained from measured
quantities, that is, the moments of the $Q_n$ distribution: our method
relies only on measurable quantities, and therefore can easily be
implemented.

\section*{Acknowledgements}
N.~B.\ acknowledges the support of the ``Actions de Recherche
Concert\'ees'' of ``Communaut\'e Fran\c{c}aise de Belgique'' and
IISN--Belgium.

\end{document}